\documentclass[aps,showpacs,superscriptadress,preprint]{revtex4}
\usepackage{graphicx}
\usepackage{amsmath}
\usepackage{amssymb}
\usepackage{mathrsfs}
\usepackage{stmaryrd}
\usepackage{amsthm}
\usepackage{hyperref}
\hypersetup{hypertex=true,
	colorlinks=true,
	linkcolor=red,
	anchorcolor=green,
	citecolor=blue}
\usepackage{caption}
\usepackage[titletoc]{appendix}

\usepackage{tikz,xcolor,hyperref}

\definecolor{lime}{HTML}{A6CE39}
\DeclareRobustCommand{\orcidicon}{
\begin{tikzpicture}
\draw[lime, fill=lime] (0,0)
circle[radius=0.16]
node[white]{{\fontfamily{qag}\selectfont \tiny \.{I}D}}; 
\end{tikzpicture}
\hspace{-3mm}
}
\foreach \x in {A, ..., Z}{%
\expandafter\xdef\csname orcid\x\endcsname{\noexpand\href{https://orcid.org/\csname orcidauthor\x\endcsname}{\noexpand\orcidicon}}
}

\begin{document}
\title{Employing shadow radius to constrain extra dimensions in black string space-time with dark matter halo}

\author{
Zening Yan$^{1}$\footnote{Corresponding author: z.n.yan.bhtpr@gmail.com or znyan21@m.fudan.edu.cn}\hspace{-2mm}\orcidA{}
} 

\affiliation{
\small 1. Department of Physics \& Center for Field Theory and Particle Physics \& Center for Astronomy and Astrophysics, Fudan University, No.2005 Songhu Road, Yangpu District, Shanghai 200438, China\\
}

%\date{December 20, 2024}

\begin{abstract}
We study the shadow of five-dimensional black strings immersed in dark matter environment.
We show the influence of the momentum characteristic parameter $\widetilde{P_{y}}$ of the extra compact dimension on the shadow radius, and we provide the constraint range $0~\leq~\widetilde{P_{y}}~\lesssim~0.171215$ in the static Schwarzschild black string solution based on the observed data.
Furthermore, we give the effective range $2.16138~\text{mm}~\lesssim~\ell~<~2.6~\text{mm}$ for the extra length $\ell$ and $\mathrm{k}~\lesssim~0.07398$ for the compactness parameter $\mathrm{k}$ in the environment near a black hole.
We also find that the momentum characteristic parameter $\widetilde{P_{y}}$ of extra dimension will affect the effective range of dark matter related parameters when using shadow radius as a constraint tool.
\end{abstract}


\maketitle

\tableofcontents

\section{Introduction}
The so-called shadow of a black hole is a dark area of shadow formed on the plane of celestial sphere under the action of a strong gravitational field, which is surrounded by an emission ring.
The attention of research on black hole shadows has increased significantly because scientists have made a series of breakthrough achievements in the field of black hole observational imaging in the last few years.
In 2000, Falcke et al. first proposed the observability of shadow images of black holes \cite{Falcke:1999pj}.
In 2019,  the Event Horizon Telescope (EHT) collaboration institution unveiled the first observational images of the supermassive black hole $\mathrm{M87}^{\star}$ \cite{EventHorizonTelescope:2019dse}.
In 2022, images of the supermassive black hole Sagittarius $\mathrm{A}^{\star}$ ($\mathrm{Sgr~A}^{\star}$) at the center of the Milky Way were also released \cite{EventHorizonTelescope:2022wkp}.
Afterwards, the EHT successively displayed polarized views of $\mathrm{M87}^{\star}$ \cite{EventHorizonTelescope:2021bee} and $\mathrm{Sgr~A}^{\star}$ \cite{EventHorizonTelescope:2024rju} related to the magnetic field around them.
In April 2023, an international team of scientists led by Ru-Sen Lu gave a panoramic image of black hole shadows and jets in the Messier 87 (M87) galaxy by using the Global Millimeter VLBI Array (GMVA), the Atacama Large Millimeter/submillimeter Array (ALMA) and the Greenland Telescope (GLT) \cite{Lu:2023bbn}.
Also in April 2023, Lia Medeiros et al. announced their results that they have developed a new principal-component interferometric modeling (PRIMO) technique that can reconstruct EHT image data to produce a sharper shadow image of the black hole \cite{Medeiros:2023pns}.
In August 2024, the EHT achieved the highest diffraction-limited angular resolution on Earth by utilizing the 345 $\text{GHz}$ frequency, providing clearer shadow images of black holes \cite{EventHorizonTelescope:2024xos}. 
It is expected that the new results will make the image clearer by 50\%.
In addition, EHT provided a series of effective ranges for shadow radii through observations of black holes $\mathrm{M87}^{\star}$ and $\mathrm{Sgr~A}^{\star}$, and applied these ranges to reasonably constrain physical quantities in space-time such as Reissner-Nordstr\"{o}m (RN), Kerr, Bardeen, Hayward, Frolov, Janis-Newman-Winicour (JNW), Kazakov-Solodhukin (KS) and Einstein-Maxwell-dilaton (EMd-1) \cite{EventHorizonTelescope:2021dqv, EventHorizonTelescope:2020qrl, EventHorizonTelescope:2022xqj}.
After that, the characteristic parameters of many other black holes are further constrained \cite{Khodadi:2022pqh, Vagnozzi:2022moj, Yan:2023pxj, Uniyal:2022vdu, Pantig:2022qak}.

Randall and Sundrum proposed a gravity theory with extra dimensions that does not use the Kaluza-Klein (KK) mechanism, also known as the Randall-Sundrum (RS) braneworld model \cite{Randall:1999ee}.
This model can be interpreted as embedding a (3+1)-dimensional submanifold (i.e., brane) in the five-dimensional Anti-de Sitter (AdS) space-time background (i.e., bulk), in which the elementary particles (except the graviton) are tightly confined to the brane, and the extra dimension exhibits $\mathbb{Z}_2$ symmetry.
Moreover, Gregory-Laflamme instability \cite{Gregory:1993vy} indicates that the black string is unstable near the AdS horizon, and the way to overcome it is to stay away from the AdS horizon.
The horospherical coordinates of the two RS models constructed with five-dimensional AdS space-time as the background are
\begin{equation}\label{eq:etay}
\mathrm{d}s^2=e^{\frac{-2|y|}{L}}\eta_{i j} \mathrm{d}x^i \mathrm{d}x^j+\mathrm{d}y^2, 
\end{equation}
where $\eta_{i j}$ is the four-dimensional Minkowski metric and $L$ is the AdS radius.
If the Minkowski metric in Eq. \eqref{eq:etay} is replaced by any Ricci flat metric, then the Einstein equations are still satisfied \cite{Chamblin:1999by, Maartens:2010ar}. 
Therefore, the natural choice is to rewrite $\eta_{i j}$ as a Ricci flat black hole solution:
\begin{equation}
\mathrm{d}s^2=e^{\frac{-2|y|}{L}}\mathfrak{g}_{\mu\nu} \mathrm{d}x^\mu \mathrm{d}x^\nu+\mathrm{d}y^2,
\end{equation}
where $\mathfrak{g}_{\mu\nu}$ is an arbitrary four-dimensional Einstein vacuum solution.
Here, the metric transformation $\mathfrak{g}_{\mu\nu} = e^{{2|y|}/{L}} g_{\mu\nu}$ can be selected, or $L\rightarrow\infty$ can be taken directly (i.e. $\mathfrak{g}_{\mu\nu} = g_{\mu\nu}$), then the metric can be expressed as 
\begin{equation}\label{eq:le}
\mathrm{d}s^2 = {g}_{\mu\nu} \mathrm{d}x^\mu \mathrm{d}x^\nu+\mathrm{d}y^2 = -h \left(\mathrm{d}t\right)^2+\frac{1}{f} \left(\mathrm{d}r\right)^2+r^2 \left(\mathrm{d}\Omega\right)^2 + \left(\mathrm{d}y\right)^2,
\end{equation}
where $\left(\mathrm{d}\Omega\right)^2$ is given by
\begin{equation}
\left(\mathrm{d}\Omega\right)^2=\left(\mathrm{d}\theta_1\right)^2+\sin^2\theta_1 \left(\mathrm{d}\theta_2\right)^2.
\end{equation}
If the metric on the four-dimensional brane is the Schwarzschild solution, which is $h=f=\left(1-2M_\mathrm{BH}\right)/r$, then this solution is the Schwarzschild black string solution \cite{Grunau:2013oca, Maartens:2010ar}.
It can be interpreted as the surface of each constant $y$ corresponds to a four-dimensional static Schwarzschild solution, and the black string solution has a linear singularity along all $y$.
In the braneworld, taking into account the five-dimensional graviton effect, the vacuum outside of the spherical star is generally not the Schwarzschild space-time but the brane solution of a static localized black hole with a five-dimensional gravitational correction, which has been discovered \cite{Germani:2001du, Kanti:2001cj, Visser:2002vg, Kanti:2003uv}, while the corresponding bulk metric has not been found. 
In this work, I replaced ${g}_{\mu\nu}$ with the asymptotically flat black hole solution.
Although this will bring some inevitable problems, this attempt is necessary before the exact solution of the bulk metric is given.

Many astronomical observations suggest that dark matter may be concentrated near the supermassive black hole at the center of the galaxy \cite{Bertone:2004pz}.
This dark matter near the black hole will profoundly change the geometric structure of space-time, turning it into a metric with anisotropic fluid ``hair'', also known as a black hole solution with a dark matter ``halo''.
Different dark matter density distributions correspond to different space-time geometries \cite{Zhang:2021bdr}.
This paper selects a static solution given by Cardoso et al., which describes a black hole immersed in a dark matter halo with a Hernquist-type density distribution at the center of a galaxy \cite{Cardoso:2021wlq}.
In addition, another space-time geometry immersed in the perfect fluid of dark matter is selected for comparison.

This paper is organized as follows:
In Section $\mathrm{\uppercase\expandafter{\romannumeral2}}$, I briefly introduced black string space-time geometry and elaborated on two metrics associated with dark matter.
In Section $\mathrm{\uppercase\expandafter{\romannumeral3}}$, I gave a detailed derivation of the equation for calculating the shadow radius in the black string space-time background.
In Section $\mathrm{\uppercase\expandafter{\romannumeral4}}$, I showed all the numerical results and provided the constraint ranges for multiple physical quantities by utilizing the observational data of the shadow radius.
In Section $\mathrm{\uppercase\expandafter{\romannumeral5}}$, I provided a brief summary of the entire paper.

Throughout the paper, the metric signature $(-,+,+,+,+)$ is selected, and a system of units $G=c=1$ is used.

\section{The geometry of space-time}

\subsection{The geometry of the black string at the center of a galaxy}
The fluid-hairy solution at the center of the galaxy can be interpreted as a black hole (black sting) immersed in a dark matter halo with anisotropic fluid. \cite{Cardoso:2021wlq, Stuchlik:2021gwg, Konoplya:2021ube}
This exact solution shows that the matter distribution is compatible with the Hernquist-type density distribution
\begin{equation}
\rho_{\text{H}}=\frac{\mathfrak{M} a_0}{2\pi r\left(r+a_0\right)^3},
\end{equation}
therefore, this model can effectively reflect the actual distribution of dark matter.

The metric function $h$ and $f$ in the line element \eqref{eq:le} are given by
\begin{equation}
h=\left(1-\frac{2M_{\mathrm{BH}}}r\right) \exp\left[-\pi\sqrt{\frac{\mathfrak{M}}{\xi}}+2\sqrt{\frac{\mathfrak{M}}{\xi}}\arctan\left(\frac{r+a_0-\mathfrak{M}}{\sqrt{\mathfrak{M} \xi}}\right)\right],
\end{equation}
\begin{equation}
f=1-\frac{2m(r)}{r},
\end{equation}
where the mass distribution of characteristic matter around the black string is 
\begin{equation}
m(r)=M_{\mathrm{BH}}+\frac{\mathfrak{M} r^2}{\left(a_0+r\right)^2}\left(1-\frac{2M_{\mathrm{BH}}}{r}\right)^2,
\end{equation}
and
\begin{equation}
\xi = 2a_0-\mathfrak{M}+4M_\mathrm{BH}.
\end{equation}
The parameters $\mathfrak{M}$ and $a_0$ represent the total mass of the ``halo'' and a typical length-scale characterizing the extension of the dark matter halo, respectively.

This solution corresponds to the dark matter density function
\begin{equation}
\rho=\frac{\mathfrak{M} \left(a_0 + 2M_\mathrm{BH}\right) \left(1-\frac{2M_{\mathrm{BH}}}r\right)}{2\pi r\left(r+a_0\right)^3}.
\end{equation}
It is different from the Hernquist-type density distribution due to the interaction between dark matter and the black string at the center of the galaxy.
This space-time solution has a horizon at $r=2M_\mathrm{BH}$ and a physical singularity at $r=0$.
Another position $r = \mathfrak{M} - a_{0} \pm \sqrt{\left(\mathfrak{M} - 2a_{0} - 4M_{\mathrm{BH}}\right) \mathfrak{M}}$ cannot be defined as a singularity because $\xi > 0$ must be satisfied.
The ADM mass can be obtained by
\begin{equation}
\lim_{r\to\infty} m(r) = \mathfrak{M} + M_{\mathrm{BH}}.
\end{equation}
In addition, $M_{\mathrm{BH}} \ll \mathfrak{M} \ll a_{0}$ should be satisfied in the actual situation of astrophysics.

For convenience, this metric will be named ``$\uppercase\expandafter{\romannumeral1}$'' in this paper.

\subsection{The geometry of a black string immersed in the perfect fluid of dark matter}
The black hole solution with a perfect fluid of dark matter is obtained when considering the gravity theory minimally coupled
with gauge field \cite{Li:2012zx, Atamurotov:2021hck}
\begin{equation}
\mathcal{S}=\int dx^4\sqrt{-g}\left(\frac{1}{16\pi G}\mathcal{R}+\frac{1}{4}F^{\mu\nu}F_{\mu\nu}+\mathcal{L}_{\text{DM}}\right),
\end{equation}
where $G$ is the Newton gravity constant, $F_{\mu\nu}$ is the Maxwell field strength, $\mathcal{L}_{\text{DM}}$ is the Lagrangian density for the perfect fluid of dark matter. 
After variation, the field equation is obtained as follows
\begin{equation}
\begin{aligned}
\mathcal{R}_{\mu\nu}-\frac{1}{2}g_{\mu\nu}\mathcal{R} &= -8\pi G \Big[ T_{\mu\nu}^{(\text{OM})}+T_{\mu\nu}^{(\text{DM})} \Big], \\
F^{\mu\nu}{}_{;\nu} &= 0, \\
F^{\mu\nu;\alpha}+F^{\nu\alpha;\mu}+F^{\alpha\mu;\nu} &= 0, \\
\end{aligned}
\end{equation}
where $T_{\mu\nu}^{(\text{OM})}$ is the energy-momentum tensor for ordinary matter and $T_{\mu\nu}^{(\text{DM})}$ is the energy-momentum tensor for the perfect fluid of dark matter.

The metric function $h$ and $f$ in the line element \eqref{eq:le} are given by
\begin{equation}
h \equiv f=1-\frac{2M_{\mathrm{BH}}}{r}+\frac{Q^2}{r^2}+\frac{\gamma}{r}\ln\left(\frac{r}{|\gamma|}\right),
\end{equation}
where $Q$ is the charge of the black string and $\gamma$ is the intensity parameter that describes the perfect fluid of dark matter.
And its singularity is located at $r=0$.

For convenience, this metric is named ``$\uppercase\expandafter{\romannumeral2}$'' in this paper.

\section{Geodetic equation and geometric shape of shadows}
In this section, I will provide a detailed derivation of the formulas required to calculate black string shadows.
Readers can also refer to the papers \cite{Grunau:2013oca, Junior:2021svb, Tang:2022bcm}.

\subsection{Null geodesics}
The Lagrangian equation for an uncharged test particle with mass $m$ is given as follows
\begin{equation}
\widetilde{\mathcal{L}}=\frac{1}{2} g_{\mu\nu} \dot{x}^{\mu}\dot{x}^{\nu} = \frac{1}{2} g_{\mu\nu} \frac{\mathrm{d}x^{\mu}}{\mathrm{d}\tau}\frac{\mathrm{d}x^{\nu}}{\mathrm{d}\tau},
\end{equation}
where the black dot is expressed as the derivative with respect to the affine parameter $\tau$.
The momentum equation of the test particle is
\begin{equation}
P_\mu=\frac{\partial\widetilde{\mathcal{L}}}{\partial\dot{x}^\mu}=g_{\mu\nu}\dot{x}^\nu=\dot{x}_\mu,
\end{equation}
and its components are calculated as follows:
\begin{equation}\label{eq:pmu}
\begin{aligned}
&P_{t}=\frac{\partial\widetilde{\mathcal{L}}}{\partial\dot{t}}= \Big(-h\Big) \dot{t},  \\
&P_{r}=\frac{\partial\widetilde{\mathcal{L}}}{\partial\dot{r}}=\left(\frac{1}{f}\right) \dot{r},  \\
&P_{{\theta}_{1}}=\frac{\partial\widetilde{\mathcal{L}}}{\partial\dot{\theta}_{1}}=\Big({r}^{2}\Big) \dot{\theta}_{1},  \\
&P_{{\theta}_{2}}=\frac{\partial\widetilde{\mathcal{L}}}{\partial\dot{\theta}_{2}}=\Big({r}^{2} \sin^2\theta_{1} \Big) \dot{\theta}_{2},  \\
&P_{y}=\frac{\partial\widetilde{\mathcal{L}}}{\partial\dot{y}}= \dot{y}.  \\ 
\end{aligned}
\end{equation}

Next, the $\mathscr{H}$ expression is given by
\begin{equation}
\mathscr{H}=P_\mu\dot{x}^\nu-\widetilde{\mathcal{L}}=\frac{1}{2}\left( g_{\mu\nu} \right)\dot{x}^\mu\dot{x}^\nu=-\frac{1}{2}{m}^2,
\end{equation}
and then in order to separate the variables of the geodesic equation, I will employ the Hamilton-Jacobi method
\begin{equation}\label{eq:HJ}
\frac{\partial S}{\partial\tau}=-\mathscr{H}=-\frac{1}{2} \left( g^{\mu\nu} \right)\cdot\frac{\partial S}{\partial x^\mu}\frac{\partial S}{\partial x^\nu},
\end{equation}
where $S$ is the Jacobi action of the test particle. 
Rewrite the above equation into the expression for each component as follows:
\begin{equation}\label{eq:HJc}
\begin{aligned}
-2\left( \frac{\partial S_\tau}{\partial\tau} \right) =&  -\frac{1}{h} \cdot\left(\frac{\partial S_t}{\partial t}\right)^2 + f\cdot\left(\frac{\partial S_r}{\partial r}\right)^2  \\
&+\frac{1}{{r}^{2}} \cdot\left(\frac{\partial S_{\theta_1}}{\partial\theta_1}\right)^2 + \frac{1}{{r}^{2}\sin^2\theta_{1}} \cdot\left(\frac{\partial S_{\theta_2}}{\partial\theta_2}\right)^2  \\
&+\left(\frac{\partial S_y}{\partial y}\right)^2  \\
\end{aligned}.
\end{equation}
The Jacobi ansatz action is expressed as a separable solution
\begin{equation}
S=\frac{1}{2} m^2\tau-\mathsf{E} t+S_r\left( r \right) + S_{\theta_{1}} \left( \theta_{1} \right)+\mathsf{L} \phi+P_{y} y,
\end{equation}
each component in the Jacobi action satisfies
\begin{equation}\label{eq:SC}
\begin{aligned}
&\frac{\partial S_{\tau}}{\partial \tau}=\frac{1}{2}{m}^2, \\
&\frac{\partial S_{t}}{\partial t}=P_t=\frac{\partial\widetilde{\mathcal{L}}}{\partial\dot{t}}=\Big(-h\Big) \dot{t} \equiv -\mathsf{E}, \\
&\frac{\partial S_{r}}{\partial r}=P_r=\frac{\partial\widetilde{\mathcal{L}}}{\partial\dot{r}}=\left(\frac{1}{f}\right) \dot{r}, \\
&\frac{\partial S_{\theta_1}}{\partial\theta_1}=P_{{\theta}_{1}}=\frac{\partial\widetilde{\mathcal{L}}}{\partial\dot{\theta}_{1}}=\Big({r}^{2}\Big) \dot{\theta}_{1}, \\
&\frac{\partial S_{\theta_2}}{\partial\theta_2}=P_{\theta_2}=\frac{\partial\widetilde{\mathcal{L}}}{\partial\dot{\theta}_{2}}=\Big({r}^{2} \sin^2\theta_{1} \Big) \dot{\theta}_{2} \equiv \mathsf{L}, \\
&\frac{\partial S_{y}}{\partial y}=P_{y}=\frac{\partial\widetilde{\mathcal{L}}}{\partial\dot{y}}= \dot{y}, \\
\end{aligned}
\end{equation}
where the conserved quantities $\mathsf{E}$, $\mathsf{L}$ and $P_{y}$ are energy, angular momentum and momentum, respectively.
$P_{y}$ is the constant derived from the extra compact dimension $y$.
And then, Eq. \eqref{eq:SC} is substituted into Eq. \eqref{eq:HJc} to obtain
\begin{equation}
\begin{aligned}
0=& \Bigg\{ \frac{{r}^{2}}{h} {\mathsf{E}}^2 - \frac{{r}^{2}}{f}\left(\dot{r}\right)^2 - {\mathsf{L}}^2 - {r}^{2} \left(P_{y}\right)^2 - {m}^2 {r}^{2} \Bigg\} - \Bigg\{ {r}^{4}\left(\dot{\theta}_{1}\right)^2 + \left( \cot^{2}\theta_{1} \right){\mathsf{L}}^2 \Bigg\}\\
\end{aligned}.
\end{equation}
Utilizing the Carter constant $\mathscr{K}$ constant \cite{Carter:1968ks}, the above equation can be separated into
\begin{equation}
\begin{aligned}
&\mathscr{K} = \Bigg\{ \frac{{r}^{2}}{h} {\mathsf{E}}^2 - \frac{{r}^{2}}{f}\left(\dot{r}\right)^2 - {\mathsf{L}}^2 - {r}^{2} \left(P_{y}\right)^2 - {m}^2 {r}^{2} \Bigg\}, \\
&\mathscr{K} = \Bigg\{ {r}^{4}\left(\dot{\theta}_{1}\right)^2 + \left( \cot^{2}\theta_{1} \right){\mathsf{L}}^2 \Bigg\}. \\
\end{aligned}
\end{equation}
Subsequently, I derived the first-order differential equations for the geodesics of the test particles, which yields
\begin{equation}\label{eq:geo}
\begin{aligned}
\dot{t}&=\frac{\mathsf{E}}{h}, \\
\dot{r}&=\pm\sqrt{\frac{f}{h} {\mathsf{E}}^2 - \frac{f}{r^{2}} \left(\mathscr{K}+{\mathsf{L}}^2\right) - f \left[\left(P_{y}\right)^2+{m}^2\right]}, \\
\dot{\theta}_{1}&=  \pm\sqrt{\frac{\mathscr{K} - \left( \cot^{2}\theta_{1} \right){\mathsf{L}}^2}{{r}^{4}}}, \\
\dot{\theta}_{2}&=
\frac{\mathsf{L} }{{r}^{2} \sin^2\theta_{1}}, \\
\dot{y}&= P_{y}. \\
\end{aligned}
\end{equation}
It can be seen that the momentum of the test particle in the extra dimension appears in the form of $\left(P_{y}\right)^2$ and contributes to the mass, that is, the effective mass of the test particle is written as $\sqrt{\left(P_{y}\right)^2+{m}^2}$.
When $m=0$, the above geodesic equations correspond to the equations of motion for a photon.
Therefore, when considering the motion of a photon in space-time with an extra dimension $y$, it is completely equivalent to the motion of a massive particle with mass $m=P_{y}$ in space-time without an extra dimension.

The shadow boundary of the black string can be determined by establishing an effective potential $V_{\text{eff}}$, which depends on the unstable circular orbits of the photons. 
The radial equation of photon motion is as follows
\begin{equation}
\left(\dot{r}\right)^2+V_{\text{eff}}=0,
\end{equation}
therefore,
\begin{equation}
V_{\text{eff}}=-\left(\dot{r}\right)^2=-\frac{f}{h} {\mathsf{E}}^2 + \frac{f}{r^{2}} \left(\mathscr{K}+{\mathsf{L}}^2\right) + f \left(P_{y}\right)^2.
\end{equation}
Introducing the characteristic parameters
\begin{equation}
\begin{aligned}
{\zeta}&=\frac{\mathsf{L}}{\mathsf{E}}, \\
\eta&=\frac{\mathscr{K}}{{\mathsf{E}}^2}, \\
\widetilde{P_{y}}&=\frac{P_{y}}{\mathsf{E}}. \\
\end{aligned}
\end{equation}
These characteristic parameters can describe the photon properties near the black string.
After that, the effective potential is rewritten as
\begin{equation}\label{eq:veff}
V_{\text{eff}} = \Bigg\{ -\frac{f}{h} + \frac{f}{r^{2}}\cdot\left(\eta + {\zeta}^2\right) + f \cdot \left(\widetilde{P_{y}}\right)^2 \Bigg\} {\mathsf{E}}^2.   
\end{equation}
The relationship between the characteristic parameters in the effective potential and the shadow of the black string will be elaborated in a later subsection.

\subsection{Celestial coordinates}
The celestial coordinates $\left(\mathrm{X}, \mathrm{Y}\right)$ \cite{Kuang:2022ojj, Abdujabbarov:2016hnw} will be employed to define the shape of shadow boundary observed by an observer at infinity,
\begin{equation}\label{eq:XY}
\begin{aligned}
&\mathrm{X}=\lim_{\mathring{r}\to\infty}\left[ \left( \left. -r^{2}\sin{\theta}_{1}\frac{d{\theta}_{2}}{dr} \right)\right|_{\left(\mathring{r}, \mathring{\theta}_{1}\right)} \right], \\
&\mathrm{Y}=\pm\lim_{\mathring{r}\to\infty}\left[ \left( \left. r^{2}\frac{d{\theta}_{1}}{dr} \right)\right|_{\left(\mathring{r}, \mathring{\theta}_{1}\right)} \right],
\end{aligned}
\end{equation}
where $\left(\mathring{r}, \mathring{\theta}_{1}\right)$ denotes the position of the observer in Boyer-Lindquist coordinates.
According to Eq. \eqref{eq:geo}, I can get
\begin{equation}\label{eq:th1th2}
\begin{aligned}
&\frac{d{\theta}_{2}}{dr}=\frac{d{\theta}_{2}}{d\tau}\cdot\frac{d\tau}{dr}=\left(\dot{\theta}_{2}\right) \left(\frac{1}{\dot{r}}\right)=\frac{\mathsf{L} \csc^{2}{\theta}_{1}}{r^2 \sqrt{\frac{f}{h} {\mathsf{E}}^2 - \frac{f}{r^{2}} \left(\mathscr{K}+{\mathsf{L}}^2\right) - f \left(P_{y}\right)^2}}, \\
&\frac{d{\theta}_{1}}{dr}=\frac{d{\theta}_{1}}{d\tau}\cdot\frac{d\tau}{dr}=\left(\dot{\theta}_{1}\right) \left(\frac{1}{\dot{r}}\right)=\frac{1}{r^2}\sqrt{\frac{\mathscr{K} - {\mathsf{L}}^2\cot^{2}\theta_{1}}{\frac{f}{h} {\mathsf{E}}^2 - \frac{f}{r^{2}} \left(\mathscr{K}+{\mathsf{L}}^2\right) - f \left(P_{y}\right)^2}}. \\
\end{aligned}
\end{equation}
After substitution, I rewrite Eq. \eqref{eq:XY} as
\begin{equation}
\begin{aligned}
&\mathrm{X}=\lim_{\mathring{r}\to\infty}\left(-\frac{\zeta \csc\mathring{\theta}_{1}}{\sqrt{\frac{f\left(\mathring{r}\right)}{h\left(\mathring{r}\right)} - \frac{f\left(\mathring{r}\right)}{\mathring{r}^{2}} \left(\eta+{\zeta}^2\right) - f\left(\mathring{r}\right) \left(\widetilde{P_{y}}\right)^2}}\right), \\
&\mathrm{Y}=\pm\lim_{\mathring{r}\to\infty}\left( \sqrt{\frac{\eta - {\zeta}^2\cot^{2}\mathring{\theta}_{1}}{\frac{f\left(\mathring{r}\right)}{h\left(\mathring{r}\right)} - \frac{f\left(\mathring{r}\right)}{\mathring{r}^{2}} \left(\eta+{\zeta}^2\right) - f\left(\mathring{r}\right) \left(\widetilde{P_{y}}\right)^2}} \right). \\
\end{aligned}
\end{equation}
Placing the observer on the equatorial hyperplane $\left(\mathring{\theta}_{1}=\frac{\pi}{2}\right)$, the celestial coordinates can be reduced to
\begin{equation}
\begin{aligned}
&\mathrm{X}=-\frac{\zeta}{\sqrt{1 - \left(\widetilde{P_{y}}\right)^2}},  \\
&\mathrm{Y}=\pm\sqrt{\frac{\eta}{1 - \left(\widetilde{P_{y}}\right)^2}}.  \\
\end{aligned}
\end{equation}

Based on the above equation, the shadow radius can be read as
\begin{equation}\label{eq:rs}
\left(R_s\right)^2 \equiv \mathrm{X}^2 + \mathrm{Y}^2 = \frac{{\zeta}^2 + \eta}{1 - \left(\widetilde{P_{y}}\right)^2}.
\end{equation}
Therefore, Eq. \eqref{eq:rs} can be used to further rewrite the effective potential \eqref{eq:veff} as
\begin{equation}
V_{\text{eff}} = \Bigg\{ -\frac{f}{h} + \frac{f}{r^{2}} \cdot \left(R_s\right)^2 \left[1 - \left(\widetilde{P_{y}}\right)^2\right] + f \cdot \left(\widetilde{P_{y}}\right)^2 \Bigg\} {\mathsf{E}}^2.  
\end{equation}
The unstable circular orbit corresponds to the maximum value of the effective potential. 
Furthermore, in order to obtain the photon sphere radius $r_p$ and shadow radius $R_s$, the effective potential needs to be given the following conditions
\begin{equation}\label{eq:veffcond}
\begin{cases}
V_{\text{eff}}=0 \\
\frac{\partial V_{\text{eff}}}{\partial r}=0 \\
\frac{\partial^2 V_{\mathrm{eff}}}{\partial r^2}<0 \\
\end{cases}.
\end{equation}

\section{Results and discussion}
In this section, I will use observation data from the EHT collaboration institution on the shadows of the $\mathrm{M87}^{\star}$ and the $\mathrm{Sgr~A}^{\star}$ black holes to constrain the calculated numerical results.
Specifically, three effective ranges will be employed.
The effective range of shadow radius for $\mathrm{M87}^{\star}$ \cite{EventHorizonTelescope:2021dqv, EventHorizonTelescope:2020qrl} is
\begin{equation}\label{eq:cond1}
3\sqrt{3}(1-0.17) M_{\mathrm{BH}}~\lesssim~R_s~\lesssim~3\sqrt{3}(1+0.17) M_{\mathrm{BH}},
\end{equation}
with confidence levels of $\sim 68\%$.
The effective range of shadow radius for $\mathrm{Sgr~A}^{\star}$ \cite{EventHorizonTelescope:2022xqj} is
\begin{equation}\label{eq:cond2}
\begin{array}{l}
3\sqrt{3}(1-0.14) M_{\mathrm{BH}}~\lesssim~R_s~\lesssim~3\sqrt{3}(1+0.05) M_{\mathrm{BH}} \quad \mathrm{(Keck)}, \\
3\sqrt{3}(1-0.17) M_{\mathrm{BH}}~\lesssim~R_s~\lesssim~3\sqrt{3}(1+0.01) M_{\mathrm{BH}} \quad \mathrm{(VLTI)}, \\
\end{array}
\end{equation}
where $\mathrm{Keck}$ and $\mathrm{VLTI}$ are from two different observation instruments.
It should be noted that the above effective ranges apply to both spherical symmetric metric constraints and rotating axisymmetric constraints.

\subsection{Influence of extra dimensional momentum on shadow radius}
\begin{figure}[htbp]
\centering
\includegraphics[width=1\textwidth]{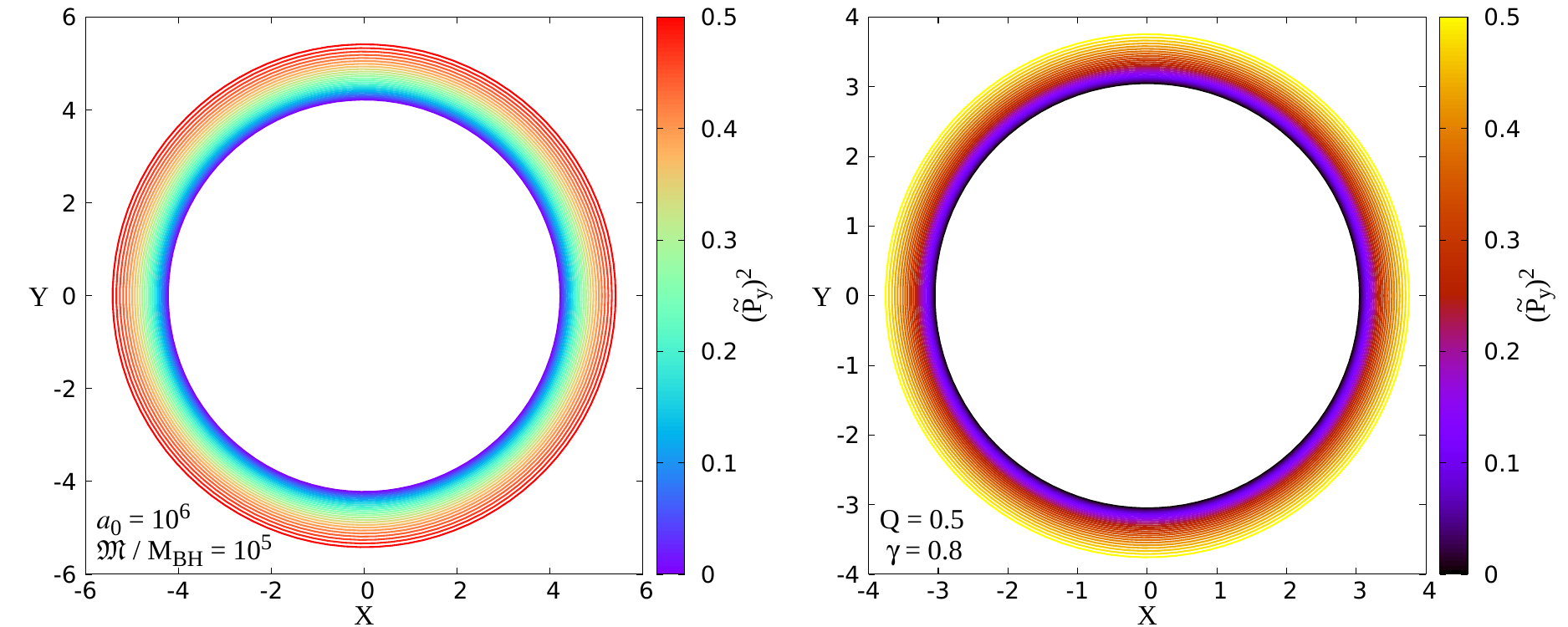}
\caption{The silhouettes corresponding to the values of different parameters $\left(\widetilde{P_{y}}\right)^2$. The left and right panels correspond to metrics $\uppercase\expandafter{\romannumeral1}$ and $\uppercase\expandafter{\romannumeral2}$, respectively. The parameter $M_\mathrm{BH} = 1$ is selected.}
\label{Fig1}
\end{figure}

\begin{figure}[htbp]
\centering
\includegraphics[width=1\textwidth]{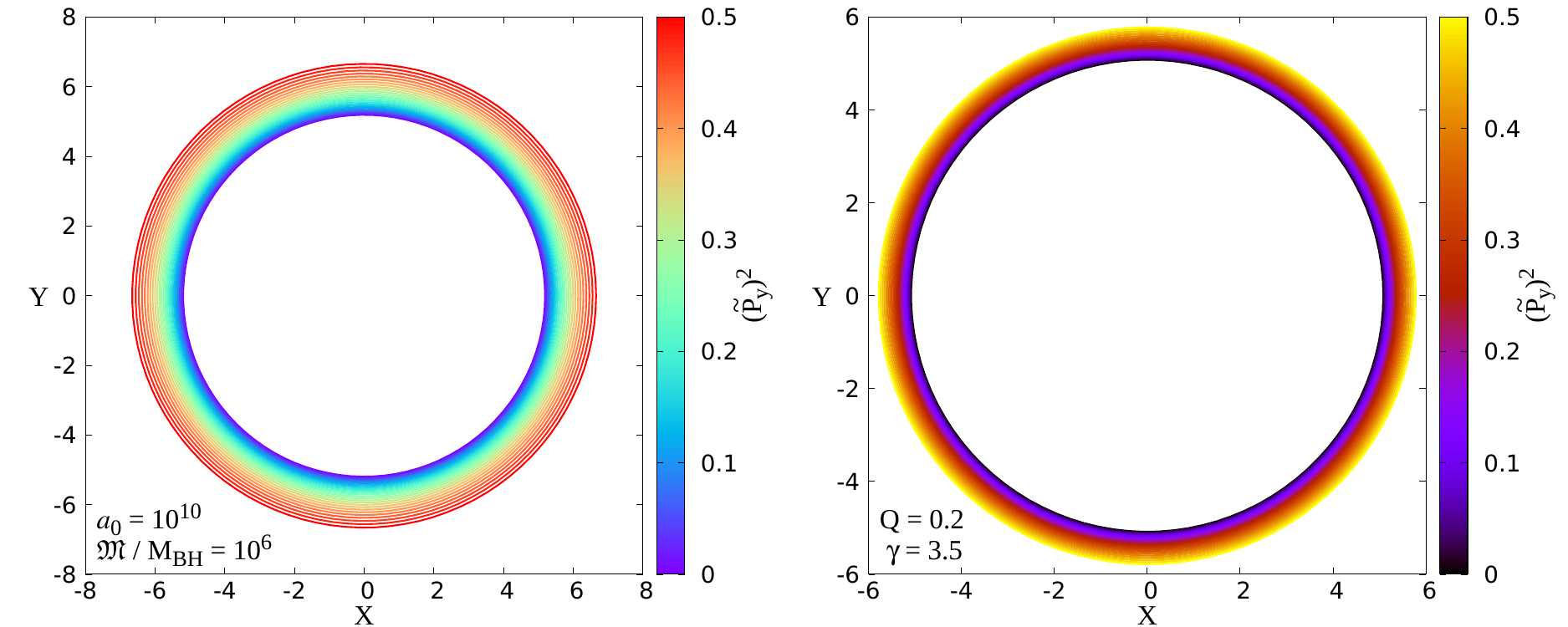}
\caption{Similar to Figs. \ref{Fig1}, but with different parameters.}
\label{Fig2}
\end{figure}

As mentioned before, the shadow radius corresponding to the momentum with extra dimensions can be obtained by solving Eq. \eqref{eq:veffcond}.
Therefore, I will provide a visual representation of the specific impact of $\widetilde{P_{y}}$ on $R_s$, as shown in Figs. \ref{Fig1} and \ref{Fig2}.
It can be clearly seen from Fig. \ref{Fig1} that an increase in $\widetilde{P_{y}}$ will lead to an increase in the shadow radius $R_s$, in other words, when the momentum characteristic parameter of the extra compact dimension increases, the shadow area also increases.

\subsection{Constraints of observation data on calculation results}
\begin{figure}[htbp]
\centering
\includegraphics[width=1\textwidth]{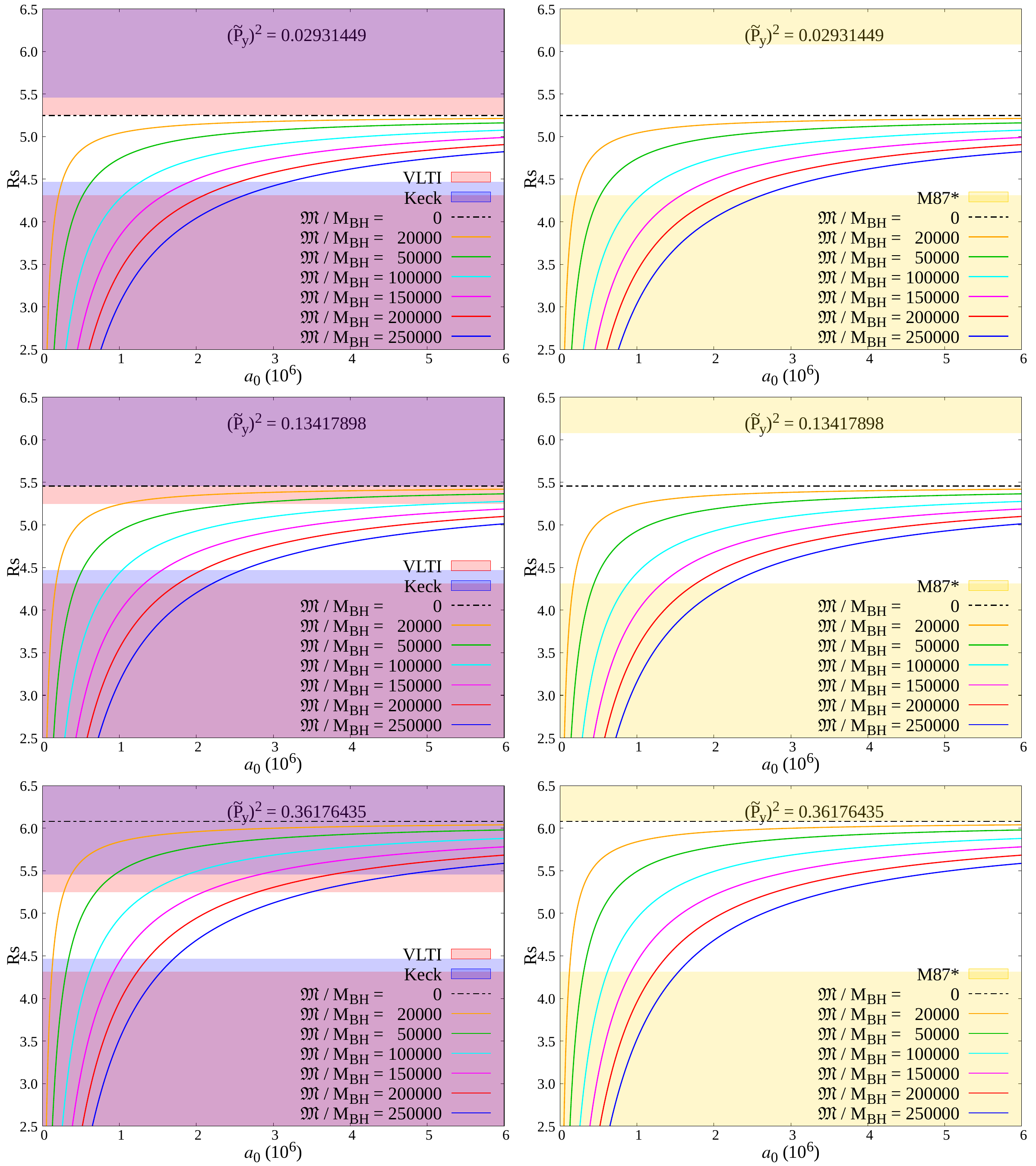}
\caption{Constraint on the shadow radius of metric $\uppercase\expandafter{\romannumeral1}$ in the black string space-time background. The three positions of the dot-dash line from top to bottom are located at the three upper boundaries of the effective range of the shadow. The parameter $M_\mathrm{BH} = 1$ is selected.}
\label{Fig3}
\end{figure}

\begin{figure}[htbp]
\centering
\includegraphics[width=1\textwidth]{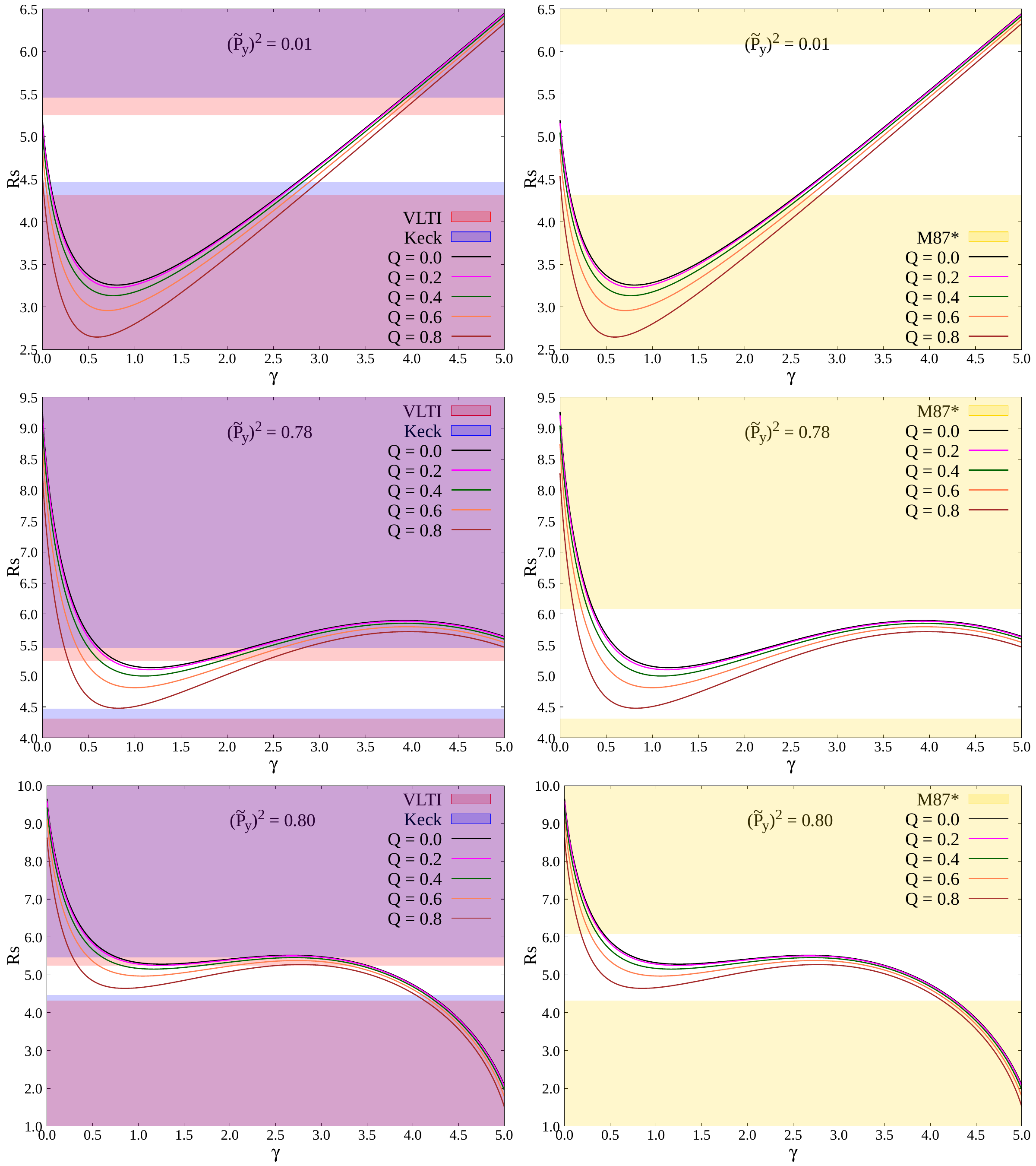}
\caption{Constraint on the shadow radius of metric $\uppercase\expandafter{\romannumeral2}$ in the black string space-time background. The parameter $M_\mathrm{BH} = 1$ is selected.}
\label{Fig4}
\end{figure}

After the conditions \eqref{eq:cond1} and \eqref{eq:cond2} are used to constrain parameter $\left(\widetilde{P_{y}}\right)^2$, the effective range of $\left(\widetilde{P_{y}}\right)^2$ can be obtained, as shown in Figs. \ref{Fig3} and \ref{Fig4}.

Before discussing the following two figures, let me briefly describe the actual values of various parameters in the galaxy.
Take the well-known Milky Way as an example, its mass is $0.9_{-0.3}^{+0.4} \times 10^{12} M_{\odot}$ \cite{Kafle:2012az}, which can be written as $\mathfrak{M}_{\mathrm{MW}} \simeq 0.9_{-0.3}^{+0.4} \times 10^{12} M_{\odot}$.
The mass of the black hole $\mathrm{Sgr~A}^{\star}$ at the center of the Milky Way is $4.297 \times 10^{6} M_{\odot}$ \cite{GRAVITY:2023avo}, denoted as $\left(M_{\mathrm{BH}}\right)_{\mathrm{Sgr~A}^{\star}} \simeq 4.297 \times 10^{6} M_{\odot}$.
The ratio can be calculated as
\begin{equation}
\frac{\mathfrak{M}_{\mathrm{MW}}}{\left(M_{\mathrm{BH}}\right)_{\mathrm{Sgr~A}^{\star}}} \simeq 0.2094 \times 10^{6}.
\end{equation}
For the $\mathrm{M87}$ galaxy, its total mass is $\mathfrak{M}_{\mathrm{M87}} \simeq (2.4\pm0.6) \times 10^{12} M_{\odot}$ \cite{Wu:2005wi}, and the mass of the supermassive black hole $\mathrm{M87}^{\star}$ at the center of the galaxy is $\left(M_{\mathrm{BH}}\right)_{\mathrm{M87}^{\star}} \simeq (6.5\pm0.2_{\text{stat}}\pm0.7_{\text{sys}}) \times 10^{9} M_{\odot}$ \cite{EventHorizonTelescope:2019ggy}. 
The ratio is
\begin{equation}
\frac{\mathfrak{M}_{\mathrm{M87}}}{\left(M_{\mathrm{BH}}\right)_{\mathrm{M87}^{\star}}} \simeq 0.3692 \times 10^{3}.
\end{equation}
Therefore, the results corresponding to the larger $\mathfrak{M}/M_{\mathrm{BH}}$ values in the figure will be shown, and the parameter $M_{\mathrm{BH}}=1$ will be fixed in this paper for calculation convenience.

First of all, Fig. \ref{Fig7} shows the constraints on the shadow radius corresponding to metric $\uppercase\expandafter{\romannumeral1}$ in the black hole space-time background, that is, the constraints when $\widetilde{P_{y}}=0$.
It can be seen from the results that the curve corresponding to each different value of $\mathfrak{M}/M_{\mathrm{BH}}$ only exists below the black dashed line ($\mathfrak{M}/M_{\mathrm{BH}}=0$), in other words, when $\mathfrak{M}/M_{\mathrm{BH}}$ increases, $R_s$ will definitely decrease.
And then, when considering the shadow constraint in the background of black string space-time, i.e., $\widetilde{P_{y}}\neq0$.
In Fig. \ref{Fig3}, all numerical results are shown to shift upwards as a whole due to the proportional relationship between $\widetilde{P_{y}}$ and $R_s$.
And different $\left(\widetilde{P_{y}}\right)^2$ values will not affect the shape of the numerical curve.
Therefore, when the black dot-dash line ($\mathfrak{M}/M_{\mathrm{BH}}=0$) coincides with the upper boundary of the effective constraint range of the shadow radius, the corresponding maximum value of $\left(\widetilde{P_{y}}\right)^2$ can be calculated, so that the true range of $\left(\widetilde{P_{y}}\right)^2$ can be determined.
The three valid ranges of $\left(\widetilde{P_{y}}\right)^2$ are as follows
\begin{equation}\label{eq:pycon}
\begin{array}{l}
0~\leq~\left(\widetilde{P_{y}}\right)^2~\lesssim~0.36176435 \quad \mathrm{M87}^{\star},\\
0~\leq~\left(\widetilde{P_{y}}\right)^2~\lesssim~0.13417898 \quad \mathrm{(Keck)},\\
0~\leq~\left(\widetilde{P_{y}}\right)^2~\lesssim~0.029314488 \quad \mathrm{(VLTI)}.\\
\end{array}
\end{equation}

When discussing metric $\uppercase\expandafter{\romannumeral2}$, the constraint of black hole space-time background can still be referred to first. 
From Figs. \ref{Fig8} and \ref{Fig9}, it can be seen that parameter $\gamma$ presents two effective regions, roughly $(0, 0.2)$ and $(3, 4)$.
Moreover, the influence of parameters $\gamma$ and $Q$ on $R_s$ can also be clearly observed through this figure.
When considering the black string space-time background corresponding to metric $\uppercase\expandafter{\romannumeral2}$, namely $\widetilde{P_{y}}\neq0$.
It can be seen from Fig. \ref{Fig4} that the effective area of the parameter after the shadow radius constraint will show a complex evolution with the change of $\left(\widetilde{P_{y}}\right)^2$ value, which implies that the change of $\left(\widetilde{P_{y}}\right)^2$ value has a significant impact on the effective range of $\gamma$.

\begin{figure}[htbp]
\centering
\includegraphics[width=0.7\textwidth]{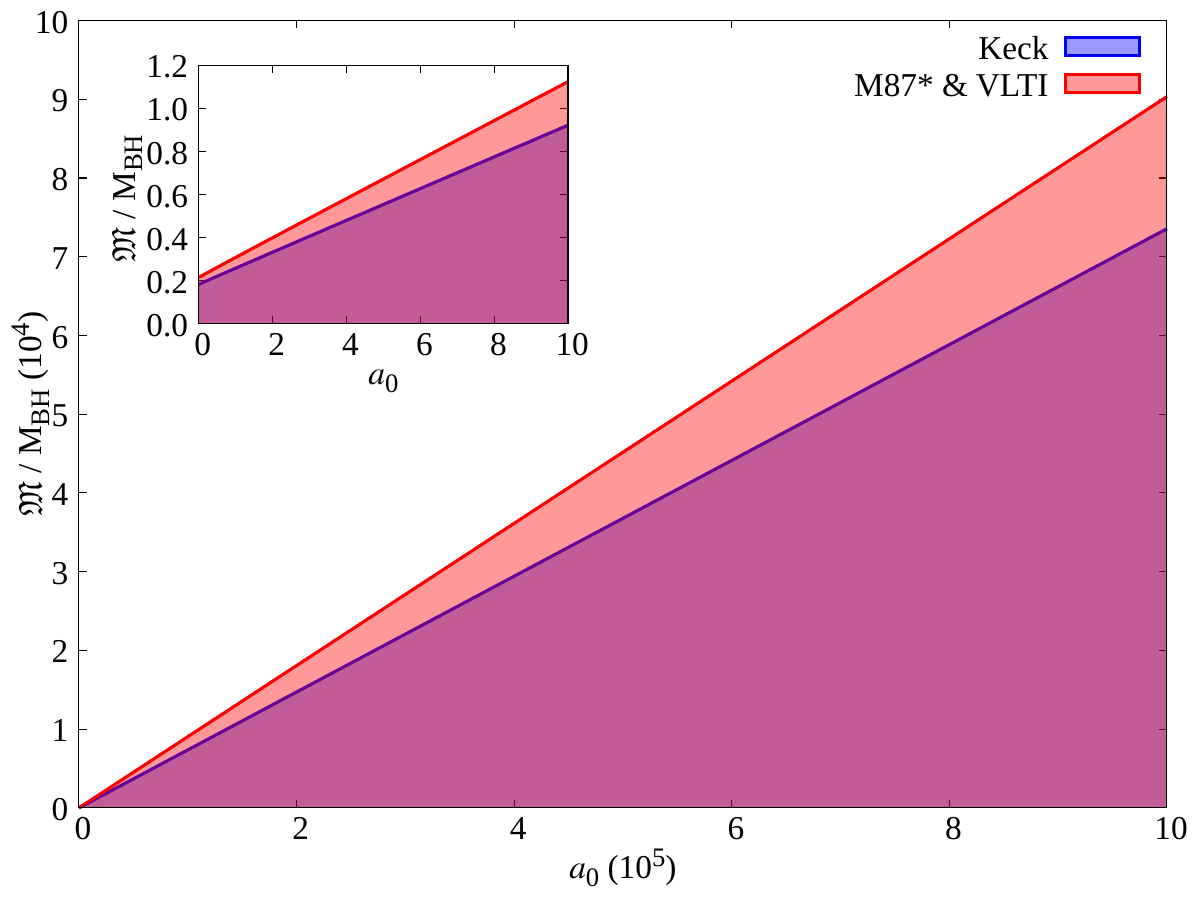}
\caption{The effective range of parameter $\mathrm{k}$ ($\mathfrak{M}/a_{0}$) in the black hole space-time background corresponding to metric $\uppercase\expandafter{\romannumeral1}$.
The red and blue areas correspond to $\mathrm{k} \lesssim 0.09097$ and $\mathrm{k} \lesssim 0.07398$, respectively.}
\label{Fig5}
\end{figure}

Let us shift our focus again to metric $\uppercase\expandafter{\romannumeral1}$ in the black hole (black string) space-time background.
In Fig. \ref{Fig7}, for each curve corresponding to different $\mathfrak{M}/M_{\mathrm{BH}}$ values, there must be an intersection point with the lower boundary of the effective range of shadow radius.
And as the parameter $\mathfrak{M}/M_{\mathrm{BH}}$ increases, the parameter $a_0$ also increases.
Therefore, the function between $\mathfrak{M}/M_{\mathrm{BH}}$ and $a_0$ can be given corresponding to the lower boundary of different shadow constraints.
It can be seen from Fig. \ref{Fig5} that the relationship between $\mathfrak{M}/M_{\mathrm{BH}}$ and $a_0$ is approximately a direct proportional function, and this function can be expressed as
\begin{equation}
\frac{\mathfrak{M}}{a_0} \simeq \mathrm{k},
\end{equation}
where $\mathrm{k}$ is the slope and $M_{\mathrm{BH}}=1$.
Furthermore, $\mathrm{k}$ is also defined as the compactness of the ``halo'' in a galaxy.
It should be noted that the boundary in Fig. \ref{Fig5} is obtained from Fig. \ref{Fig7}.
After a simple thought, it can be determined that the region below the straight line is a reasonable region that satisfies the shadow constraint range.

The observation of galaxies roughly satisfies the condition \cite{Navarro:1995iw}
\begin{equation}
\frac{\mathfrak{M}}{a_0} \lesssim 10^{-4}.
\end{equation}
If the constraint is performed under a black hole environment, then the parameter $\mathrm{k}$ is free \cite{Cardoso:2021wlq}.
Therefore, it is feasible to constrain this parameter by means of the black hole shadow.
Obviously, the constraint range of parameter $\mathrm{k}$ in the black hole environment is much larger than the range given in the galaxy environment.

\begin{figure}[htbp]
\centering
\includegraphics[width=0.7\textwidth]{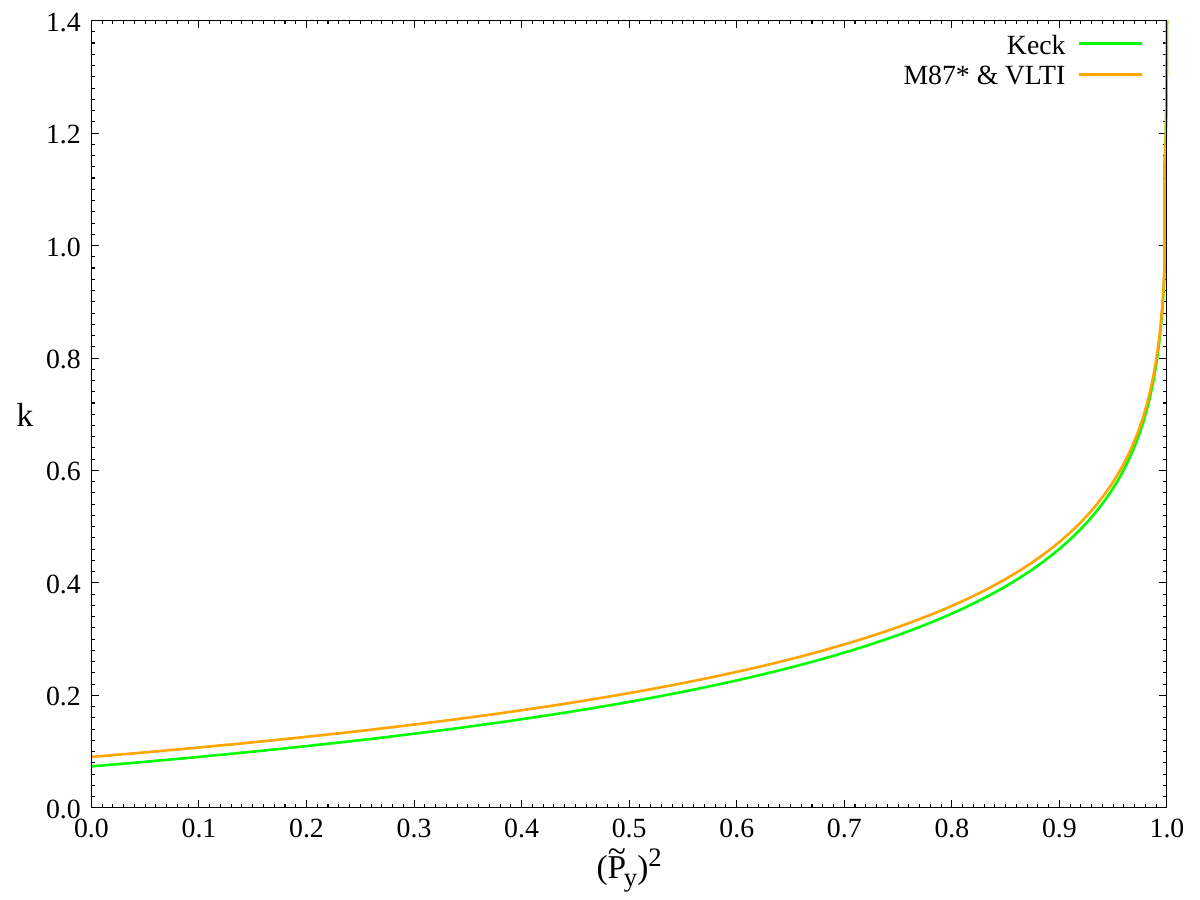}
\caption{The maximum value of parameter $\mathrm{k}$ increases as the parameter $\left(\widetilde{P_{y}}\right)^2$ increases.}
\label{Fig6}
\end{figure}

When considering the case of $\widetilde{P_{y}}\neq0$, the maximum value boundary of $\mathrm{k}$ will change. 
In Fig. \ref{Fig6}, it can be seen that the maximum value of k will increase as $\widetilde{P_{y}}$ increases.
Combined with the effective range of $\left(\widetilde{P_{y}}\right)^2$ after being constrained, it can be found that the increase of $\left(\widetilde{P_{y}}\right)^2$ has little effect on the maximum boundary $\mathrm{k}$, but only slightly increases the effective range of $\mathrm{k}$.
For the case where the constraint range is $\left(\widetilde{P_{y}}\right)^2 \lesssim 0.029314488$, the parameter $\mathrm{k}$ satisfies the condition $\mathrm{k} \lesssim 0.08$.
For the case of $\left(\widetilde{P_{y}}\right)^2 \lesssim 0.36176435$, $\mathrm{k}$ satisfies the condition $\mathrm{k} \lesssim 0.163$.

Next, the constraint on the length $\ell$ of the compact extra dimension will be given below. 
Here I will follow the idea of paper \cite{Tang:2022bcm} and use the following condition
\begin{equation}\label{eq:vyc}
\frac{v_y}{c} = \frac{c P_{y}}{\mathsf{E}} = c \widetilde{P_{y}} = \frac{2\pi\hbar n c}{\mathsf{E} \ell} = \frac{n \lambda_0}{\ell} \leq 1,
\end{equation}
where $\lambda_0$ represents the wavelength of the photons, and $n=0, \pm 1, \pm 2, \cdots$.
Based on the following inequality
\begin{equation}\label{eq:npy}
\frac{n \lambda_0}{\left(\widetilde{P_{y}}\right)_{max}} < \left(n+1\right) \lambda_0,
\end{equation}
the value of $n$ can be determined as $n=0,1$, namely $n\neq2$.
And then, according to the wavelength $\lambda_0=1.3~\text{mm}$ of the microwaves used by EHT in the observation and Eq. \eqref{eq:pycon}, constraints on the length of the extra dimensions can be obtained, where the constraint satisfying the above conditions is ${\lambda_0}/{\sqrt{0.36176435}} \lesssim \ell < 2\lambda_0$, that is, 
\begin{equation}\label{eq:lrange}
2.16138~\text{mm}~\lesssim~\ell~<~2.6~\text{mm}.
\end{equation}
Since the above constraint is obtained when $n=1$, therefore, the constraint range of $\widetilde{P_{y}}$ in terms of condition \eqref{eq:npy} should be rewritten as
\begin{equation}\label{eq:repy}
0.5~<~\widetilde{P_{y}}~\lesssim~0.601468.
\end{equation}
It should be noted that the above constraints are obtained by applying the shadow constraints of $\mathrm{M87}^{\star}$ in Eq. \eqref{eq:pycon}.
It can be seen from \eqref{eq:repy} that the constraints of $\mathrm{Sgr~A}^{\star}$ in \eqref{eq:pycon} cannot be used when selecting $n=1$.

In the paper \cite{Tang:2022bcm}, two constraints on the length of an extra dimension in the black string space-time background were also given using shadow observations of $\mathrm{M87}^{\star}$ and $\mathrm{Sgr~A}^{\star}$, which are $2.03125~\text{mm}\lesssim\ell<2.6~\text{mm}$ and $2.28070~\text{mm}\lesssim\ell<2.6~\text{mm}$ respectively.
It can be seen that the constraint range \eqref{eq:lrange} obtained in this work using $\mathrm{M87}^{\star}$ is tighter than the range given in paper \cite{Tang:2022bcm}.
However, there is no way that I can use the observations of the shadow radius of $\mathrm{Sgr~A}^{\star}$ to give a reasonable constraint on the extra dimension length $\ell$, because $\mathrm{Sgr~A}^{\star}$ gives an extremely small constraint range of $\widetilde{P_{y}}$ \eqref{eq:pycon}, so that it violates the condition \eqref{eq:npy}!

This question makes me doubt the constraint range of $\ell$ given by this method.
If we assume that the upper boundary of the effective range of the shadow radius is approximately at $\sim3\sqrt{3}(1+0.1)$, then the range of $\widetilde{P_{y}}$ is $0\leq\widetilde{P_{y}}\lesssim0.499999$.
This means that when $n=1$, the value of $\ell$ should be $2.6~\text{mm}<\ell<2.6~\text{mm}$, in other words, there is no extra dimension scale.
Referring to Fig. \ref{Fig3}, it can be seen that my assumption above is obviously reasonable,
Therefore, I believe that although the constraint range for the extra dimension length $\ell$ is given through \eqref{eq:vyc} and \eqref{eq:npy}, it does not guarantee that this range is absolutely rigorous, and we still need to be cautious about this result.

Finally, it is worth mentioning that the paper \cite{Lemos:2024wwi} has given the effective range of the $\mathrm{AdS}_5$ curvature radius of the Garriga-Tanaka black hole in the braneworld as follows:
\begin{equation}
\begin{array}{l}
L_{\left(\text{VLTI}\right)} \lesssim 2.4 \times 10^9 \text{m}, \\
L_{\left(\text{Keck}\right)} \lesssim 5.7 \times 10^9 \text{m}.
\end{array}
\end{equation}

\section{Conclusions}
We studied the shadows of two metrics associated with dark matter in the black string (black hole) space-time background.

In this work, the influence of extra dimensions on shadow radius is given, and the constraint range of the momentum characteristic parameters $\widetilde{P_{y}}$ and length $\ell$ of extra dimensions are constrained by using the observation data of shadow radius.
In addition, we have also given a constraint range for the compact parameter $\mathrm{k}$ (i.e. ${\mathfrak{M}}/{a_0}$) of the dark matter halo in the environment near the black string/hole.
Next, we focused on the preciseness of these constraints, and the conclusion is that the constraint range of parameters $\widetilde{P_{y}}$ and $\mathrm{k}$ has high confidence.
However, we should be cautious about the constraint ranges of parameters $\ell$.
Although our results are consistent with those in \cite{Tang:2022bcm}, we believe that the scope of constraints on $\ell$ still needs to be discussed in more depth.

This work also attempted to answer the following question:
Will changes in certain parameters of the extra dimension have an impact on the dark matter around the black string?
Our preliminary conclusion is that the effective range of dark matter related parameters will change with the change of the characteristic parameters of the extra dimension, that is, some parameter values that do not meet the constraint conditions may become qualified.
Moreover, this effect is also closely related to the structure of the metric.

\begin{appendices}
\section{Constraints on black hole space-time}
\begin{figure}[htbp]
\centering
\includegraphics[width=1\textwidth]{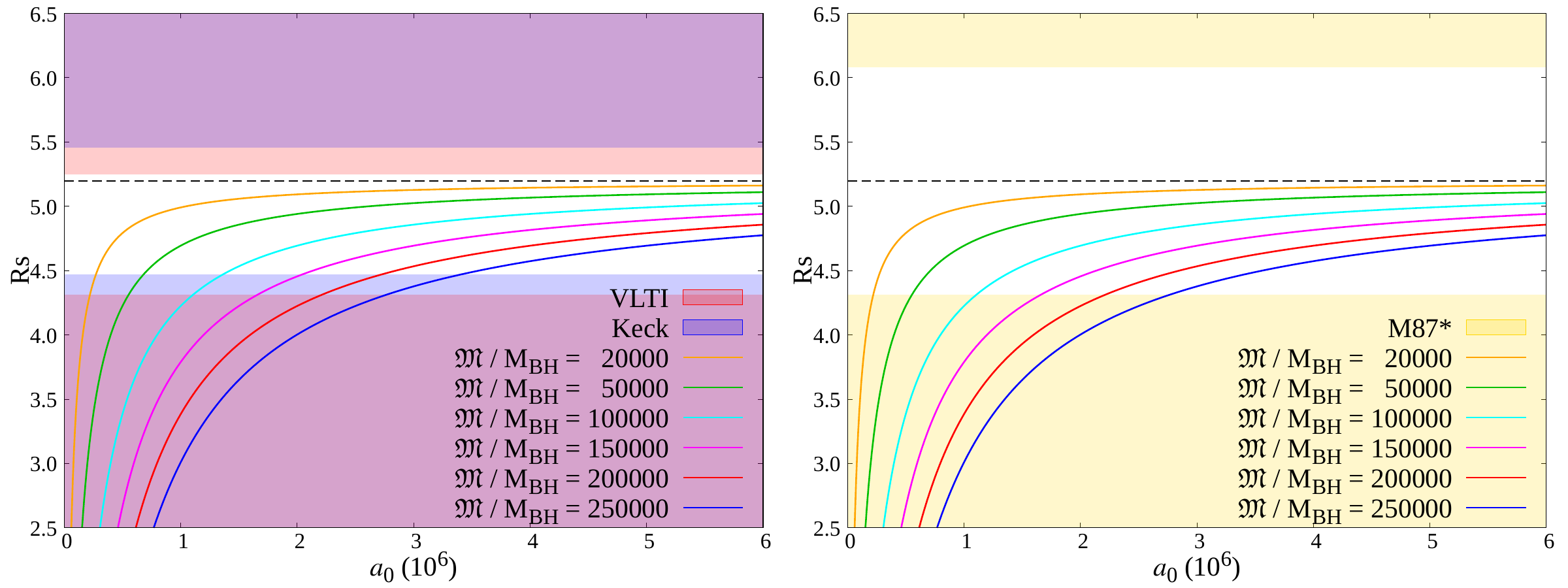}
\caption{Constraint on the shadow radius of metric $\uppercase\expandafter{\romannumeral1}$ in the black hole space-time background. The dotted line corresponds to $\mathfrak{M}=0$, which is the position of $R_s=3\sqrt{3}$.}
\label{Fig7}
\end{figure}

\begin{figure}[htbp]
\centering
\includegraphics[width=1\textwidth]{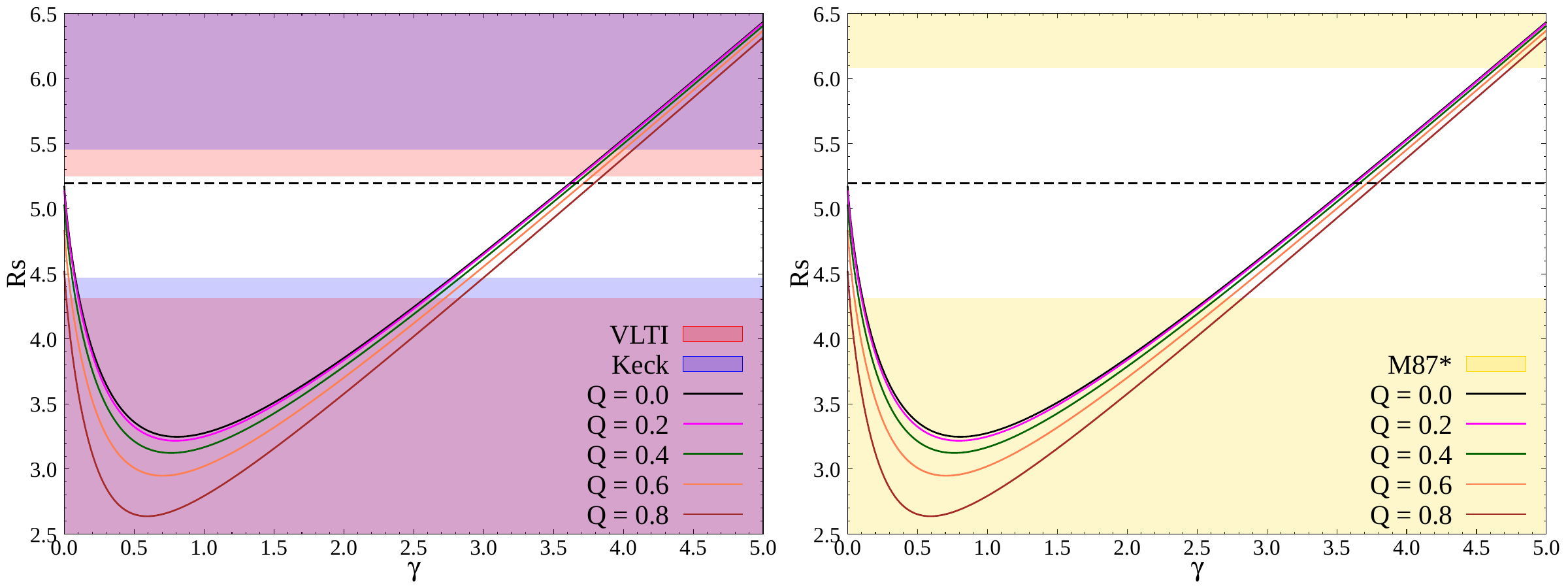}
\caption{Constraint on the shadow radius of metric $\uppercase\expandafter{\romannumeral2}$ in the black hole space-time background.}
\label{Fig8}
\end{figure}

\begin{figure}[htbp]
\centering
\includegraphics[width=1\textwidth]{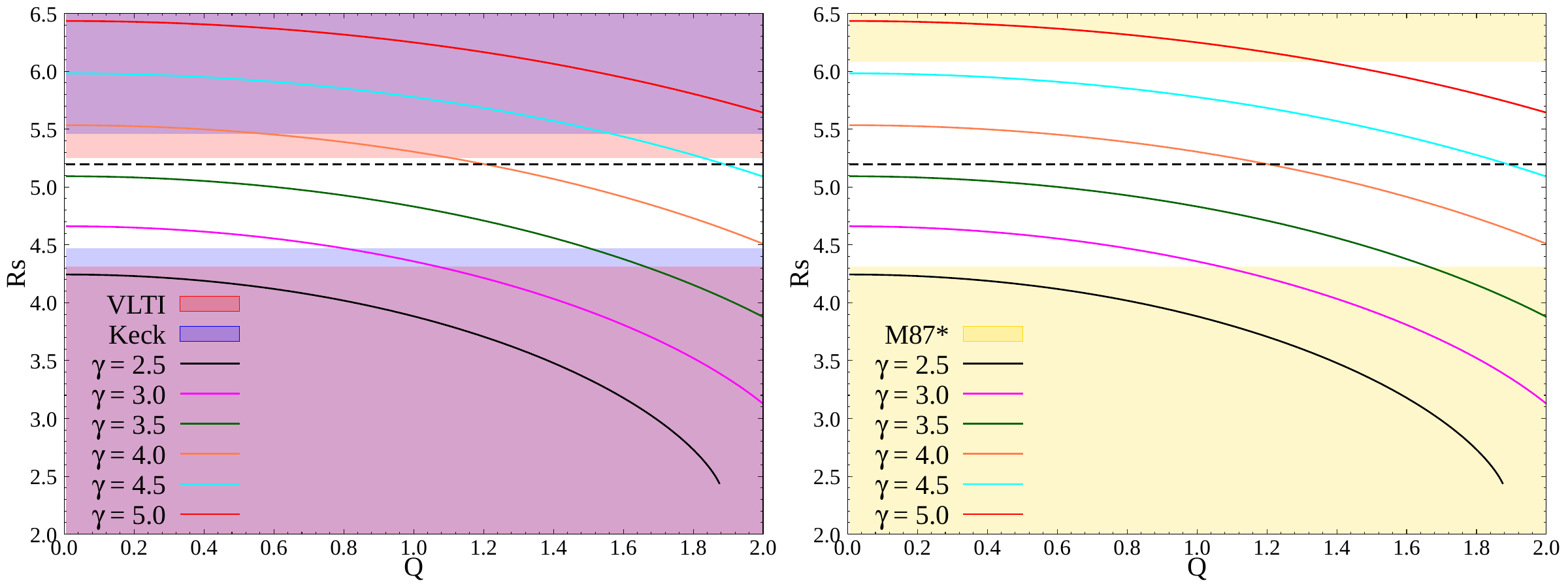}
\caption{Constraint on the shadow radius of metric $\uppercase\expandafter{\romannumeral2}$ in the black hole space-time background.}
\label{Fig9}
\end{figure}

Here, I provided constraints on the shadow radius for the metrics $\uppercase\expandafter{\romannumeral1}$ and $\uppercase\expandafter{\romannumeral2}$ in the black hole space-time background, respectively, as shown in Figs. \ref{Fig7}, \ref{Fig8} and \ref{Fig9}.

\begin{figure}[htbp]
\centering
\includegraphics[width=0.6\textwidth]{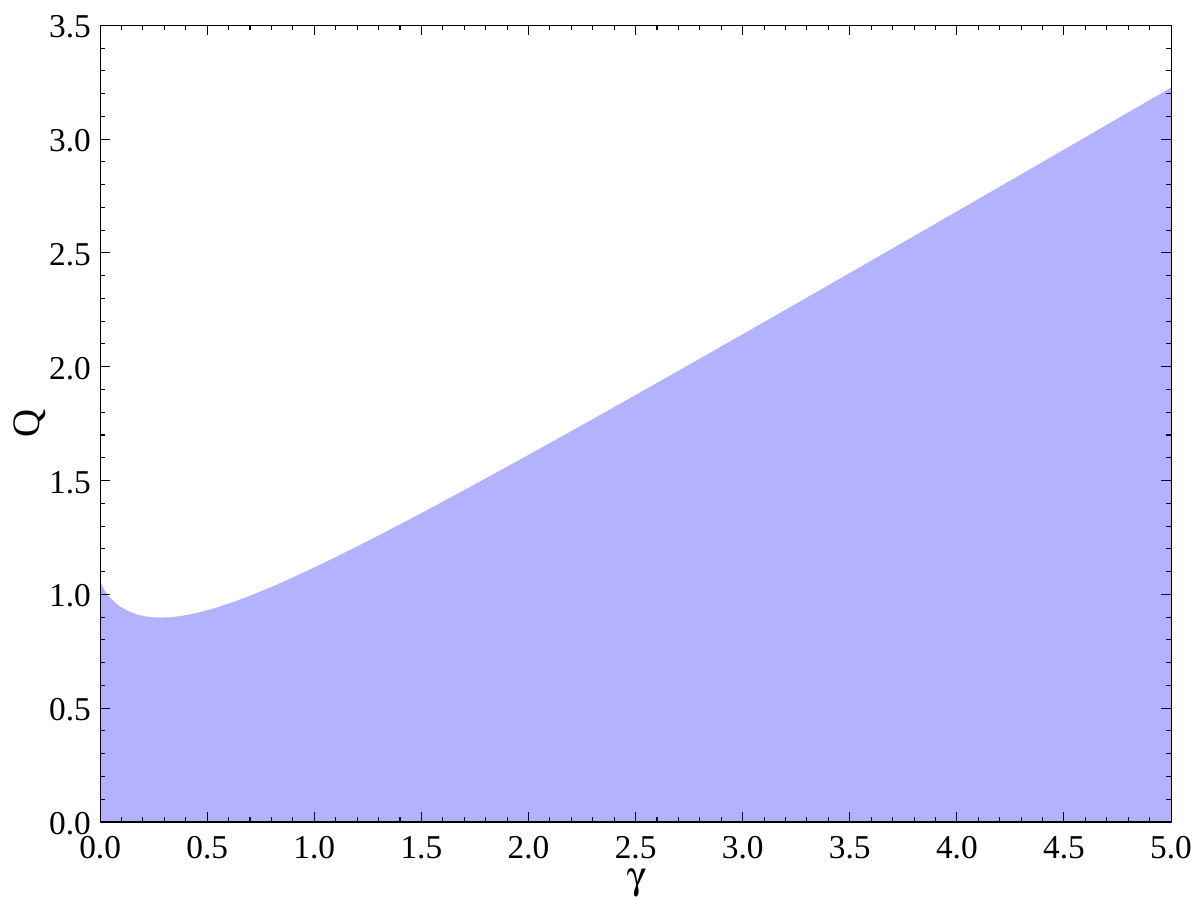}
\caption{The blue region corresponding to the existence of an event horizon in metric $\uppercase\expandafter{\romannumeral2}$.}
\label{Fig10}
\end{figure}

In addition, for the metric $\uppercase\expandafter{\romannumeral2}$, I also gave the effective regions of parameters $Q$ and $\gamma$ to ensure that no naked singularities occur in black holes, as shown in Fig. \ref{Fig10}.
In the actual calculation, $Q<0.9$ is the selected range of parameter $Q$.

\end{appendices}

\section*{Acknowledgements:}
The author acknowledges the people, teams and institutions who have contributed to this paper.

\section*{Data Availability Statement:} 
All relevant data are within the paper.

\bibliography{BHDMHED}

\end{document}